\documentclass{aastex631}

\graphicspath{{./}{figures/}}
\usepackage{amsmath,bm}
\usepackage{graphicx}

\usepackage{xcolor}
\usepackage{soul}

\begin{document}

\title{Multi-thermal jet formation triggered by flux emergence}

\author[0000-0001-8164-5633]{Xiaohong Li}
\affiliation{Centre for mathematical Plasma Astrophysics, \\
Department of Mathematics, KU Leuven, \\
Celestijnenlaan 200B, 3001, Leuven, Belgium}

\author[0000-0003-3544-2733]{Rony Keppens}
\affiliation{Centre for mathematical Plasma Astrophysics, \\
Department of Mathematics, KU Leuven, \\
Celestijnenlaan 200B, 3001, Leuven, Belgium}

\author[0000-0002-4391-393X]{Yuhao Zhou}
\affiliation{Centre for mathematical Plasma Astrophysics, \\
Department of Mathematics, KU Leuven, \\
Celestijnenlaan 200B, 3001, Leuven, Belgium}

\begin{abstract}
Flux emergence is responsible for various solar eruptions. Combining observation and simulations, we investigate the influence of flux emergence at one footpoint of an arcade on coronal rain as well as induced eruptions. The emergence changes the pressure in the loops, and the internal coronal rain all moves to the other side. The emerging flux reconnects with the overlying magnetic field, forming a current sheet and magnetic islands. The plasma is ejected outwards and heated, forming a cool jet $\sim 6000$ K and a hot X-ray jet $\sim 4$ MK simultaneously. The jet dynamical properties agree very well between observation and simulation. In the simulation, the jet also displays transverse oscillations with a period of 8 minutes, a so-called whip-like motion. The movement of the jet and dense plasmoids changes the configuration of the local magnetic field, facilitating the occurrence of Kelvin--Helmholtz instability, and vortex-like structures form at the boundary of the jet. Our simulation clearly demonstrates the effect of emergence on coronal rain, the dynamical details of reconnecting plasmoid chains, the formation of multi-thermal jets, and the cycling of cool mass between the chromosphere and the corona.

\end{abstract}


\section{Introduction} \label{sec:intro}

Magnetic flux emergence has been accepted to be essential for many solar activities. For example, the emergence of flux tubes from the convection zone into the solar atmosphere triggers the formation of sunspots and solar active regions (ARs) \citep{Zwaan1985, Zwaan1987, Cheung2010, vanD2015, ChenF2017}. Also, the interaction between preexisting field and emerging flux could lead to various eruptions, through magnetic reconnection or instabilities \citep{Heyvaerts1977, Feynman1995, ChenP2000, ChenP2011, Cheung2014, Kliem2006, JiangC2016, Toriumi2019}. Unprecedented high-resolution observations have provided us with unique opportunities to investigate flux emergence and its associated dynamics in different solar atmospheric layers, such as the triggering of spicules, H$\alpha$ surges, extreme Ultra-Violet (EUV) or X-ray jets and flares \citep{Schrijver2009, Guglielmino2010, Guglielmino2018, Schmieder2014, Wang2020, Shen2022}.  

As it is hard to study all facets of magnetic flux emergence through observations, numerical magnetohydrodynamic (MHD) simulations have been employed to mimic the emergence of non-twisted or twisted flux tubes into the solar atmosphere and their corresponding eruptions \citep{Shibata1989, Caligari1995, Archontis2004, Manchester2004, Cheung2014, FanY2001, FanY2021}. \citet{Yokoyama1995, Yokoyama1996} showed that reconnection between the emerging flux and the overlying magnetic field could drive a microflare, an H$\alpha$ surge, and an EUV jet simultaneously. \citet{Moreno2008} compared the result of a three-dimensional (3D) MHD simulation with Hinode observations and found that various properties of the jet, such as timescales and velocities, are in agreement with observations. \citet{Leake2013, leake2014} found that without the reconnection between the emerging field with the preexisting field, there would not be eruptions, and the system will develop into a quasi-static equilibrium. Clearly, magnetic reconnection is not only a consequence of the flux emergence, but also the driver of the later eruptions. Many flux emergence simulations ignore the turbulent convection in the solar interior and treat the subsurface layers as adiabatically stratified or through externally provided boundary conditions. These idealized simulations have sometimes limited treatment of the energy transfer mechanisms such as thermal conduction, radiative losses, or background heating, but allow to focus on the essential physical processes in the higher solar atmosphere \citep{Shimojo2001, Miyagoshi2004, Torok2016, Lionello2016}. Realistic simulations with a convective zone layer, that include advanced radiation transfer aspects are a more self-consistent approach to study flux emergence, which help us understand how convective motion can affect the origin and evolution of the coronal magnetic fields 
\citep{Nobrega2016,Moreno2018,Cheung2019, Hansteen2019,Nobrega2022, Toriumi2022}. Despite these advancements, realistic simulations are challenging, and hard to independently reproduce and validate, as the combined plasma and radiative physics from the solar interior to the upper layers is complex. Idealized simulations can isolate and study specific mechanisms that may be obscured in more complex realistic simulations. Here, we include all the necessary physical mechanisms without going to full sub-photosphere to corona evolutions.

In this letter, we report a solar eruption event triggered by magnetic flux emergence, where some fascinating multi-thermal aspects can be witnessed: the effect of jets on coronal rain, the guiding of both hot and cold matter along a neighbouring coronal loop system, and intricate up- and down-cycling of cool matter flows. We then present a 2.5D MHD simulation on coronal rain with emerging flux on one side of the coronal loops, to study the influence of the emerging flux on the chromospheric and coronal plasma. The simulation is idealized but fully reproducible, and includes all the important physical mechanisms such as optically thin radiative cooling, field-aligned thermal conduction, gravity and background heating. Contrary to `self-consistent' models, our heating is parametrically inserted, and not due to hyperdiffusive visco-resistive effects. Our simulation is in good agreement with the observation: we see how changes in the magnetic field guide the motion of the plasma, but also how the motion of the plasma affects the configuration of the magnetic field. This letter is organized as follows. First, we describe the details of the observations in Sect. \ref{sec:obser}, then we present the numerical setup of our simulations in Sect. \ref{sec:setup}. Sect. \ref{sec:result} list the results of our simulations, and the conclusions and discussion are given in Sect. \ref{sec:conc}.

\section{Observation} \label{sec:obser}

\begin{figure}
\includegraphics[trim = 0mm 0mm 0mm 0mm, clip, width=1.0\textwidth]{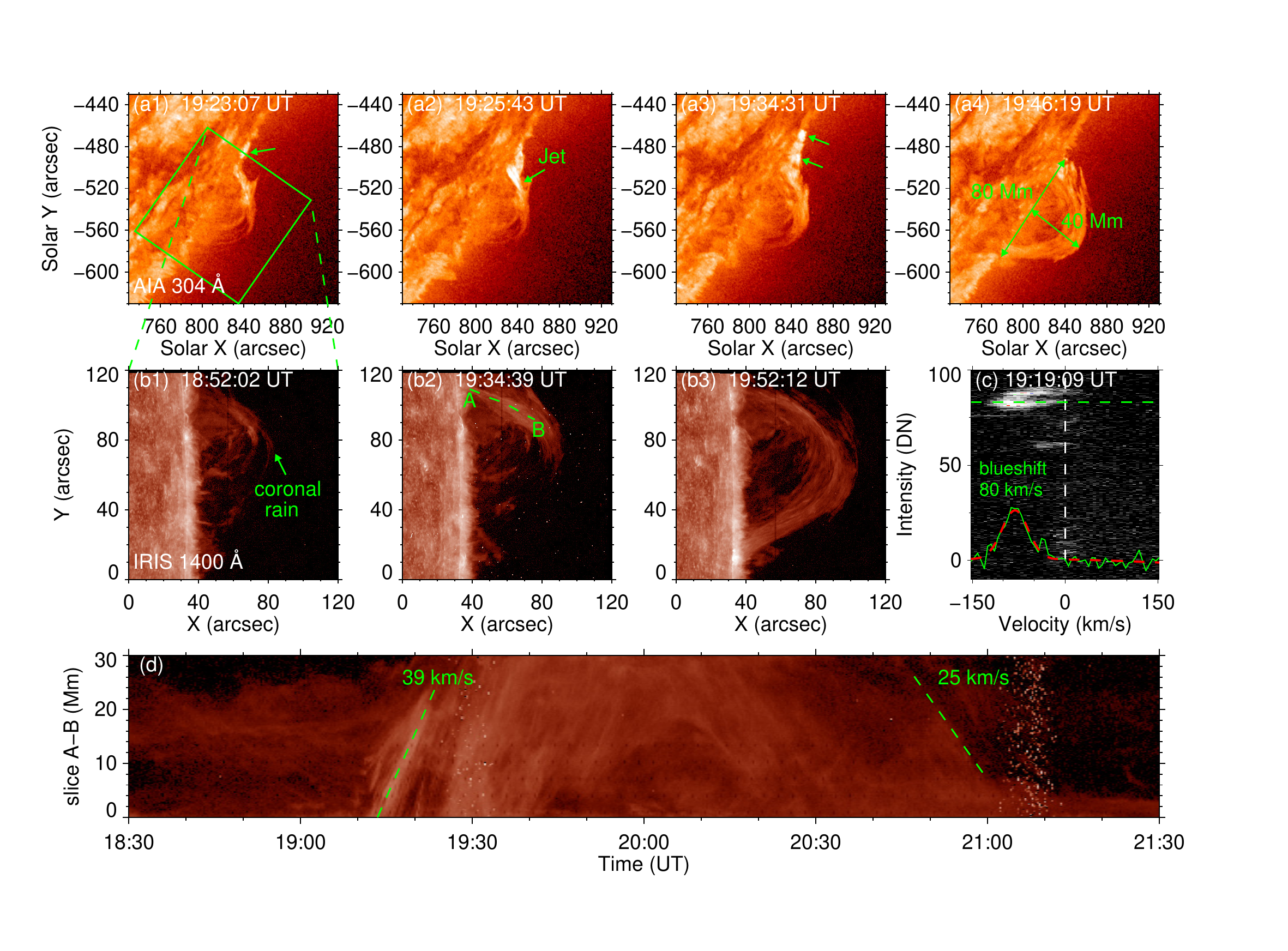}
\caption{Observations of the eruptions on 2014 October 9. Panels (a1)--(a4) and Panels (b1)--(b3) are AIA 304 {\AA} images and \emph{IRIS} 1400 {\AA} images displaying the jet, respectively. The green rectangle in panel (a1) outlines the field-of-view (FOV) of panels (b1)--(b3). The green arrows in (a3) show the plasmoid-like structures. Panel (c) shows the Si IV 1403 {\AA} spectra in the slit range of panel (b2) for Doppler velocities from $-$150 km s$^{-1}$ to 150 km s$^{-1}$.  The green solid curve is the observed profiles at the selected locations denoted by the green dashed line, and the red dashed curve is the corresponding single-Gaussian fitting. Panel (d) display the temporal evolution of flows at the position of slice A--B (see panel (b2)).
An animation (Animation1.mov) accompanying this figure is available. The upper panel of the movie is the AIA 304 {\AA} observations from 18:50 UT to 20:58 UT on October 9 2014, the cadence is 12 s and the spatial resolution is 0.$\arcsec$6. The lower movie panel is the IRIS 1400 {\AA} observations at the same time, with a resolution of 0.$\arcsec$166 pixel$^{-1}$ and a cadence of 19 s.}
\label{Fig1}
\end{figure}

An EUV jet was observed at the solar limb on 2014 October 9 by the \emph{Solar Dynamics Observatory} \citep[\emph{SDO};][]{Pesnell2012} and the \emph{Interface Region Imaging Spectrograph} \citep[\emph{IRIS};][]{DeP2014} satellites. We present the Atmospheric Imaging Assembly \citep[AIA;][]{Lemen2012} 304 {\AA} images from \emph{SDO} and \emph{IRIS} 1400 {\AA} slit-jaw-images (SJIs) of this event in Figure \ref{Fig1} (and associated movie Animation1.mov). 1400 {\AA} SJIs have a pixel size of 0.$\arcsec$166 and a cadence of 19 s, while the spatial sampling of the 304 {\AA} images is 0.$\arcsec$6 pixel$^{-1}$ and the cadence is 12 s. We also use data from the spectrograph on \emph{IRIS} in 1402.77 {\AA} to measure the Doppler shift. As shown in the accompanying animation, there is a coronal loop system at the west  limb of the solar surface, with coronal rain material inside it (see Fig. \ref{Fig1}, panels (a1) and (b1)). At around 19:10 UT, chromospheric matter is ejected upwards at the northern footpoints of the loops, and then develops into an EUV jet, denoted by the green arrows in the AIA 304 {\AA} movie (see Animation1.mov and Fig. \ref{Fig1}(a2)). From the movie, we can also see that a small bright loop emerges from the north side of the coronal loop (see the green arrow in Fig. \ref{Fig1}(a1)). Following the emergence, the jet becomes bigger and brighter, as shown in Fig. \ref{Fig1}(a2). As the emerging flux moves upwards, plasmoid structures form at the top of the interaction zone with preexisting field (indicated in Fig. \ref{Fig1}(a3)), and these are ejected outwards, becoming part of the jet. This jet flows along the loop system which is 80 Mm long, and 40 Mm high. \emph{IRIS} only captures the development of the jet, with its smaller field of view as shown by the green rectangle in panel (a1). At 19:34 UT, the jet was moving upwards, as shown in panel (b2). Analysing a slice along the position denoted by the dashed line ``A--B" (panel (b2)), we find that the upward velocity of the jet is about 39 km s$^{-1}$ (the time-distance view is provided in panel (d)). Panel (c) displays the appearance of the Si IV 1403 {\AA} spectra in the slit range of panel (b2) for Doppler velocities from -150 km s$^{-1}$ to 150 km s$^{-1}$ when the jet is moving upwards. The green curve is the observed profile along the dashed line, and the red curve is the corresponding single-Gaussian fitting. We can see that the jet has a clear Doppler blue shift with a velocity of 80 km s$^{-1}$. At the late stage of the jet, that is, after around 20:00 UT, there are backflows with velocities of about 25 km s$^{-1}$, and cool matter drops back to the northern footpoints of the loops (see panel (d)). This observation shows how a pre-existing multi-thermal loop system, containing coronal rain, gets affected by flux emergence, and reshuffles both cool and hot material along an extended neighbouring loop system. We use MHD simulations to better understand these complex plasma processes, involving eruption, reconnection, and chromospheric matter recycling.

\section{Numerical setup} \label{sec:setup}

Using the parallelized Adaptive Mesh Refinement Versatile Advection Code \citep[\href{https://amrvac.org/}{MPI-AMRVAC},][]{Keppens2012, Porth2014, Xia2018, Keppens2021}, we perform a 2.5D simulation following the setup in our previous work where we studied realistic coronal rain cycles in a randomly heated arcade system \citep{LiX2022}. To be consistent with the observations as discussed in Sect.~\ref{sec:obser}, we change our domain to a Cartesian ($x$, $y$) box with $x$ axis from $-$ 60 Mm to 60 Mm and $y$ axis from 0 Mm to 60 Mm and the effective resolution is 39 km in both directions. The system is relaxed to a quasi-equilibrium state using an exponentially decaying background heating, and then a randomized heating pattern is added to mimic the energy input from the lower solar atmosphere, as adopted in \citet{Zhou2020, LiX2022}. After about 143.1 min of localized heating, there are condensations in the loops, corresponding to the observed coronal rain inside the loops. Then we add an emerging flux region near the left footpoints of the loops, to determine its influence on the raining loops. 

For the realization of an emerging magnetic flux, we prescribe a time-dependent electrical field at the lower boundary \citep{FanY2004}. This electric field corresponds to a magnetic dipole $\bm{B_1}$ moving upwards at a constant speed $\bm{v}$:
\begin{equation}
\bm{E}= - \bm{v} \times \bm{B_1}.
\end{equation}
where $\bm{v}$ = $v_0$ $\hat{\mathbf{y}}$, and
\begin{equation}
\begin{aligned}
B_{1x} &= - B_1 \cos \left (\frac {\pi (x-x_0)}{L} \right ) \sin \theta_1 {\rm exp} \left (- \frac {\pi \sin \theta_1 (y-y_0(t))}{L}\right ),  \\
B_{1y} &= B_1 \sin \left (\frac {\pi (x-x_0)}{L} \right ) {\rm exp} \left (- \frac {\pi \sin \theta_1 (y-y_0(t))}{L}\right ),   \\
B_{1z} &= - B_1 \cos \left (\frac {\pi (x-x_0)}{L} \right ) \cos \theta_1 {\rm exp} \left (- \frac {\pi \sin \theta_1 (y-y_0(t))}{L}\right ).
\end{aligned}
\end{equation}
Here $B_1$ = 150 G and $L$ = 20 Mm.
$x_0$ and $y_0(t)$ control the instantaneous location of the emerging flux, $x_0 = -30$ Mm and we assume the center of the dipole is first located at $y_0(t=0)=y_c = - 10$ Mm, to then move up with speed $v_0$. The emerging dipole is transported into the computational domain through the time-dependent electric field at the lower boundary, and the emergence stops after around 86 min. $\theta_1$ is the angle between the emerging bipole and the $x-y$ plane. In this letter, we set $\theta_1$=0, which means the emerging flux has a 60-degree angle to the background field.

The simulation employs a three-step Runge-Kutta time integration method, with the ``vanleer" limiter and Harten-Lax-van Leer Riemann solver. We adopt the upwind constrained transport (CT) method to constrain the magnetic field divergence close to zero. We enforce symmetry of the density, total energy, $y$ component of momentum, and $B_y$, while the $x$ and $z$ components of momentum, $B_x$ and $B_z$ are forced to be anti-symmetric at the left and right boundaries. For the bottom boundary, we set the velocity to zero, except for the emergence area where the velocity $v_0$ equals 2 km s$^{-1}$. The density and pressure are fixed as predetermined in the initial condition. At the top boundary, we use a fixed zero velocity and extrapolate pressure and density according to the gravity stratification. We determine $B_x$ and $B_z$ with a zero-gradient extrapolation and derive $B_y$ using the zero-divergence constraint at both the top and bottom boundaries. Our simulation includes gravity, field-aligned thermal conduction using the Runge-Kutta Legendre super-time-stepping (RKL-STS) scheme \citep{Meyer2012, Meyer2014}, optically thin radiative losses handled with the exact integration method \citep{Townsend2009}, randomized heating, and the transition region adaptive conduction (TRAC) method \citep{Johnston2019, ZhouY2021}. This comprehensive set of physical mechanisms allows us to study the thermodynamic properties of the solar atmosphere more realistically.


\begin{figure}
\includegraphics[bb=200 150 1600 1200,clip,angle=0,scale=0.37]{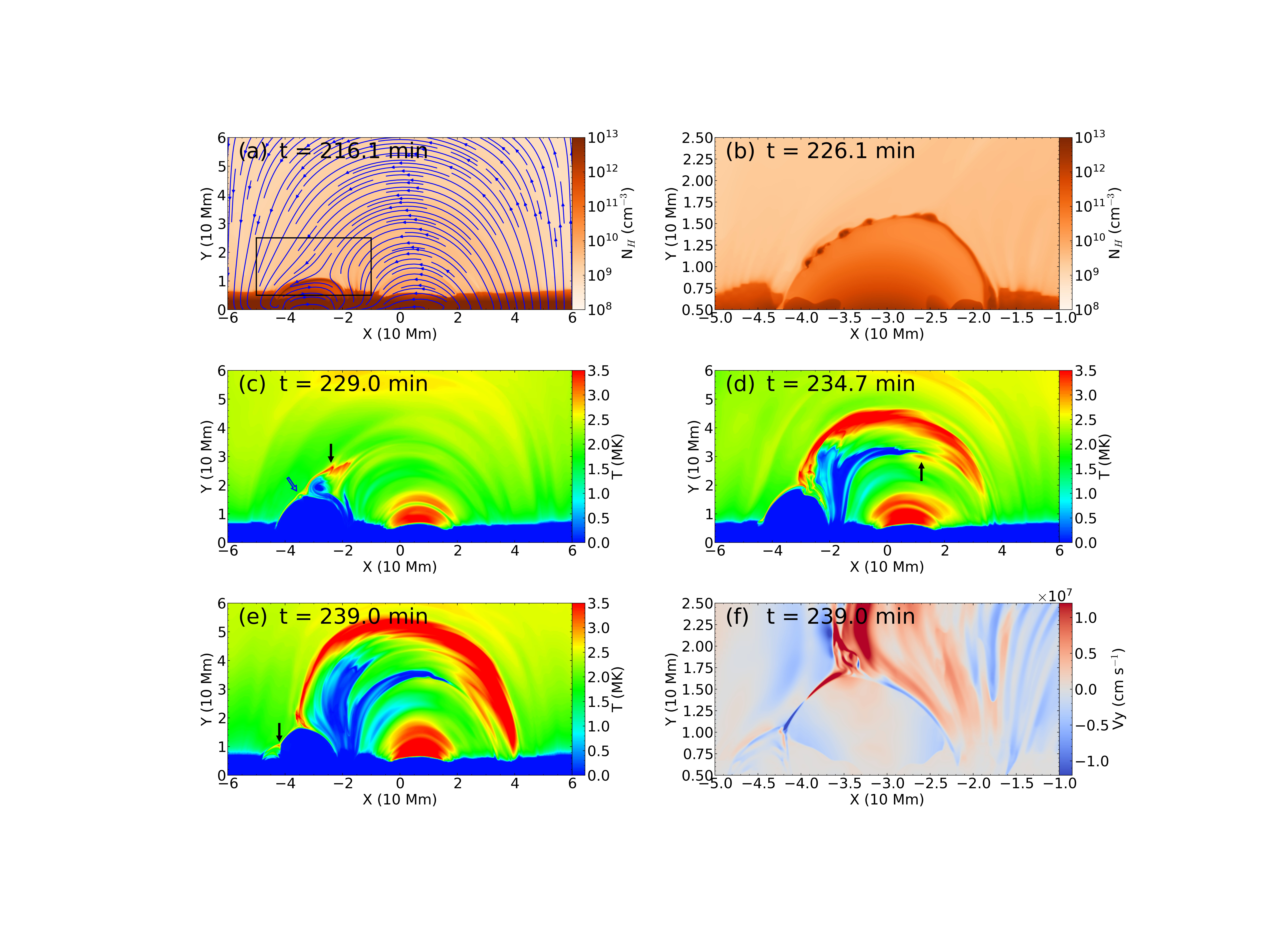}
\caption{Simulations of flux emergence and the formation of eruptions. Panels (a) and (b) are the number density maps, and the FOV of panel (b) is denoted by the black rectangle in panel (a). The blue lines in panel (a) indicate the magnetic field lines. Panels (c)--(e) illustrate the temperature change of the domain during the flux emergence, and the arrows point to localized heating areas. Panel (f) is the vertical velocity map in the zoomed-in black rectangle area as shown in panels (a)-(b).
An animation (Animation2.mov) of this figure is available, showing the evolution of number density and temperature of our simulation from $t$ = 0 min to $t$ = 300 min. The time cadence of the animation is about 1.431 min.}
\label{Fig2}
\end{figure}

\begin{figure}
\includegraphics[bb=80 0 800 400,clip,angle=0,scale=0.85]{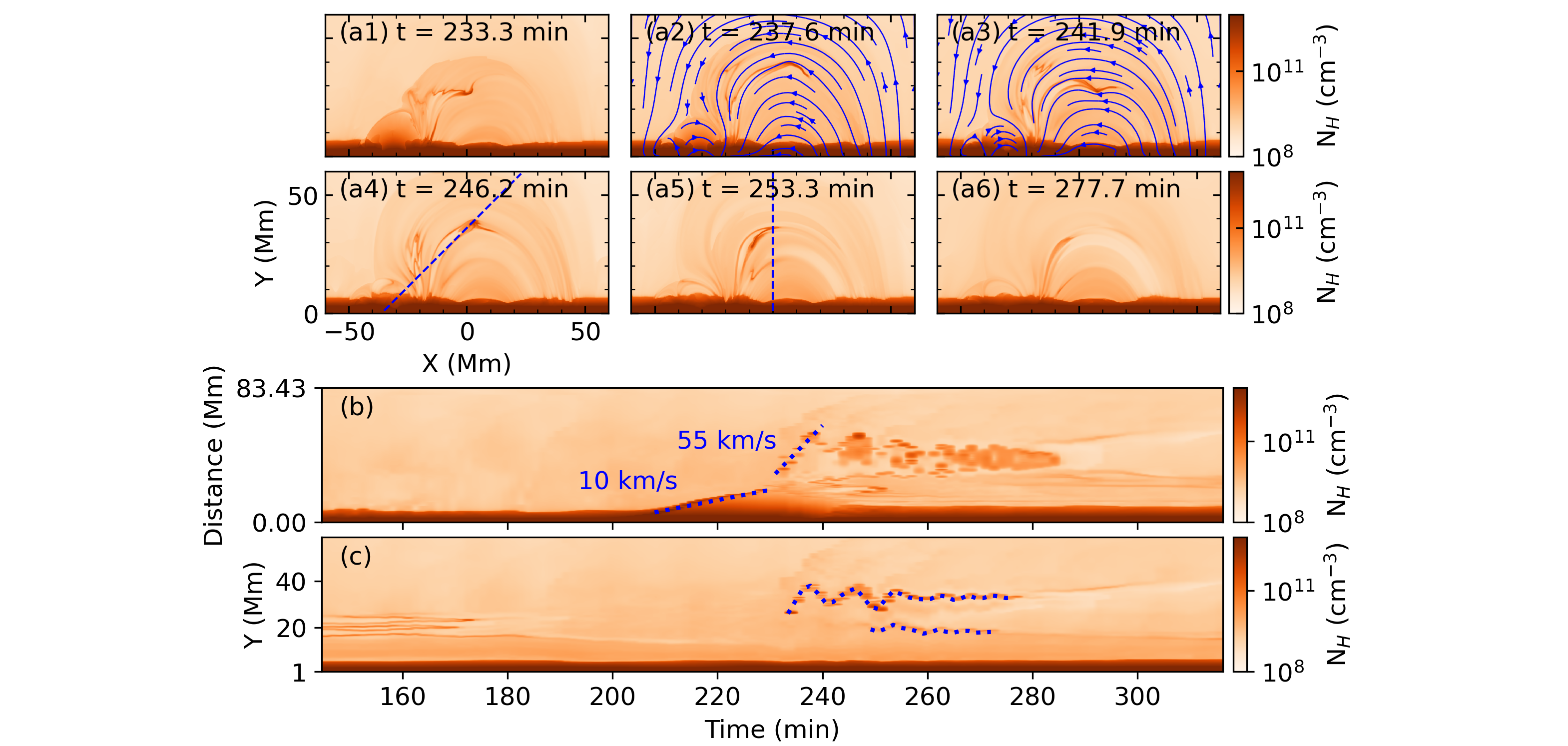}
\caption{Number density maps showing the whip-like motion of the jet. Panel (b) is the time-distance plot of the number density along the position indicated by the blue dashed line in panel (a4). Panel (c) shows the evolution of the number density along the $y$-direction as denoted by the blue dashed line in panel (a5). The blue dotted curves display the oscillations of the jet and the movement of condensations in the lower loops.}
\label{Fig3}
\end{figure}

\begin{figure}
\includegraphics[width=18cm]{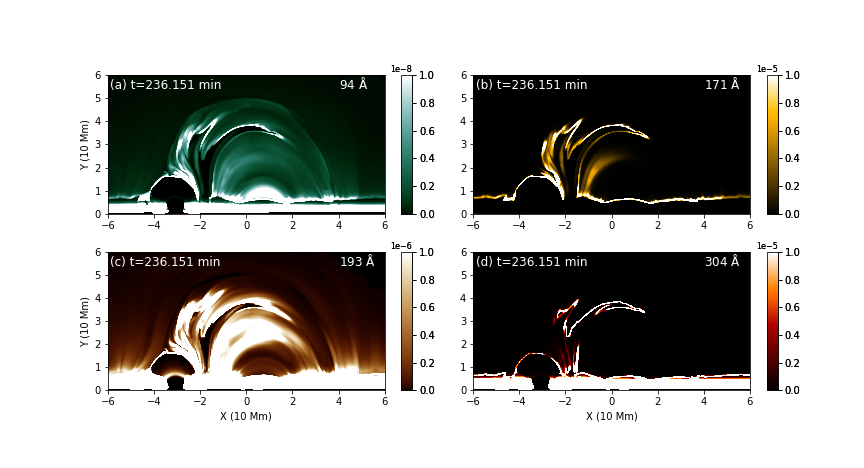}
\caption{AIA 94 {\AA}, 171 {\AA}, 193 {\AA} and 304 {\AA} synthetic images showing the eruptions at the selected time.}
\label{Fig4}
\end{figure}

\begin{figure}
\includegraphics[bb=50 100 1000 1150,clip,angle=0,scale=0.5]{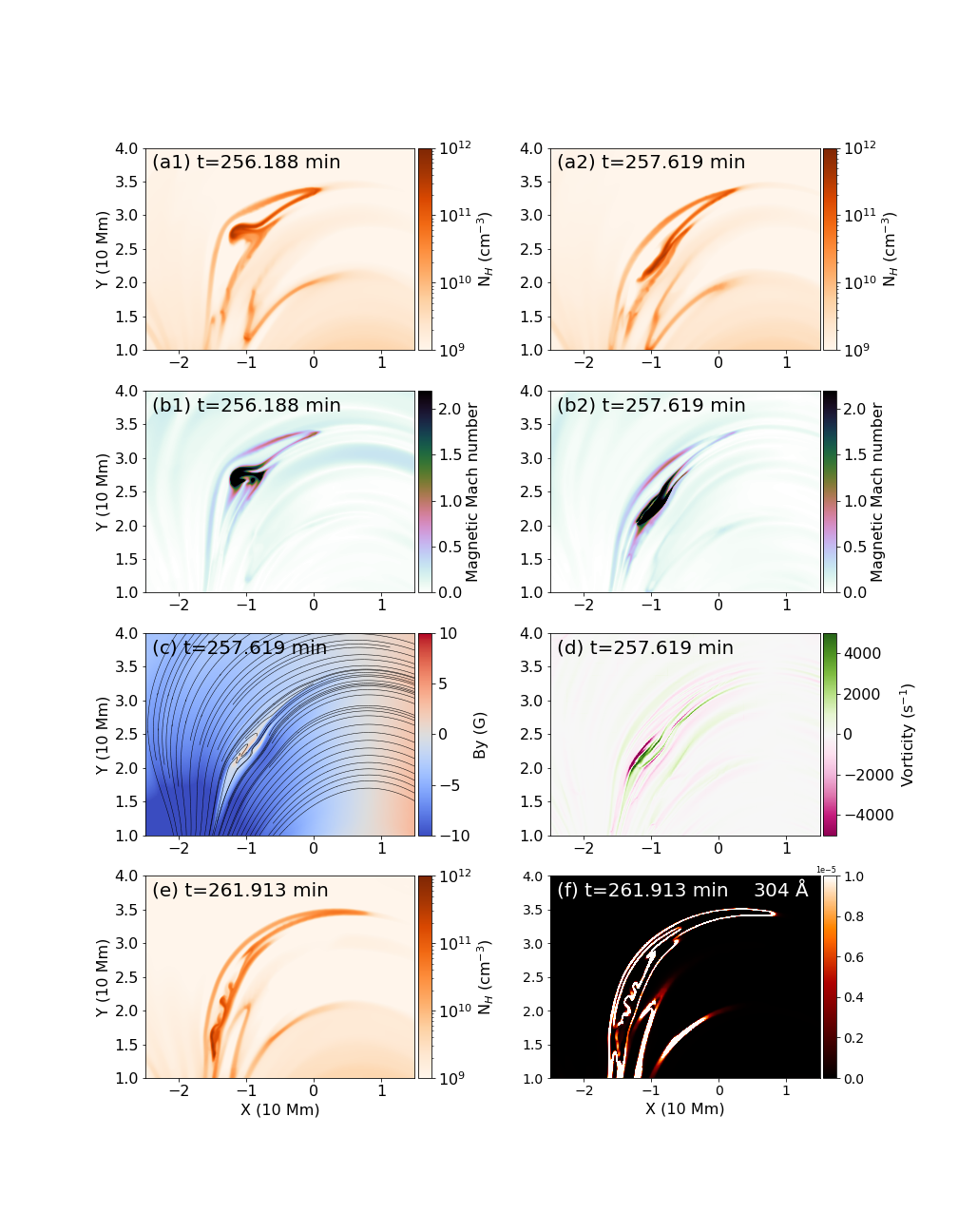}
\caption{The occurrence of Kelvin--Helmholtz instability. Panels (a1)--(a2) and panels (b1)--(b2) are the number density and Mach number maps at $t =$ 256.188 min and $t =$ 257.629 min. Panel (c) and panel (d) show the distributions of the magnetic field in the $y$ direction and the vorticity at $t =$ 257.619 min, respectively. The number density map in panel (e) and the synthetic AIA 304 {\AA} image in panel (f) display the vortex-like structures caused by the Kelvin--Helmholtz instability at $t =$ 261.913 min.}
\label{Fig5}
\end{figure}

\section{Results} \label{sec:result}

As in our previous paper \citep{LiX2022}, after turning on the localized heating, thermal instability will occur and condensations form continuously in the loops, at first demonstrating a zig-zag structure due to sympathetic runaway cooling on nested loops \citep{Fang2013,Fang2015}, as shown in Animation2.mov. Starting from this coronal rain-filled arcade system at $t =$ 144.6 min, we put an emerging flux at the left footpoint. As the emerging flux changes the pressure in the loops, the cool condensations move preferentially to the other side of the loop, instead of splitting into small blobs and dropping to both footpoints, as in the previous work \citep{Fang2013, Fang2015, LiX2022}. Figure \ref{Fig2}(a) shows the number density map and the magnetic field lines (in blue) at around 70 minutes after the emerging flux protrudes through the lower boundary. We can see that the emerging flux invades the corona with lower atmospheric plasma. After the emerging flux enters the corona, a current sheet is created on top of the emerging flux, and there is a dense loop located at the current sheet (see Animation2.mov especially between times $t = 209.0 - 224.7$ min), which is believed to be an arched filament \citep{Zwaan1985, Shibata1992}. As the dipole is emerging upwards, multiple magnetic islands form in the current sheet due to the tearing instability, as displayed in Figure \ref{Fig2}(b). Each island encloses dense and cool plasma originally in the emerging loop. The distance between each island is about 2.5 Mm. Soon after formation, the small islands merge into a large island, as shown in panel (c). Then the resulting large island is ejected horizontally out of the current sheet. At the same time, the plasma near the current sheet is heated to a high temperature, as denoted by the black arrow in panel (c). We also find hot plasmoids pointed by the lower hollow blue arrows in panel (c) and Animation2.mov at $t =$ 227.6 and 229.0 min. The temperature of the hot plasmoid is about 2.5 MK. In panels (d) and (e), the ejection of the cool plasma forms a cold H$\alpha$ surge with a temperature of 6000 K, and the nearby corresponding heating is manifested as a hot X-ray jet, whose temperature is more than 4 MK. In front of the H$\alpha$ surge, we also find some heating (see the black arrow in panel (d)), which may be caused by shock waves \citep{Forbes1983, Pontin2022}. The arrows in panel (e) and Animation2.mov between $t = 236.2 - 240.4$ min denote the heating near the transition region (TR), which could be recognized as a flare or bright point in observations. Panel (f) shows the vertical velocity, and there are clear bi-directional flows at the upper and lower edges of the current sheet, which is one of the typical manifestations of magnetic reconnection \citep{Innes1997, LiX2018}. These outflows have velocities around 400 km s$^{-1}$ (of order the local Alfv\'en speed, as expected from reconnection theory). 

The number density maps in Figure \ref{Fig3} display the dynamics of the jet. We can clearly see that the jet shows transverse oscillations, which have been termed as whip-like motions \citep{Canfield1996, Morton2012, Chandrashekhar2014}. As mentioned in the last paragraph, the emerging flux, which carries the cool plasma from the chromosphere, expands upward and reconnects with the preexisting coronal loops, forming magnetic islands and bi-directional flows. As the magnetic tension of disconnected field lines drives the plasmoid to move outwards, as it experiences the sling-shot effect, the magnetic field line gets stretched and bent due to the heavy plasma load, as shown in Fig.~\ref{Fig3}, panel (a1). As the jet shows a whip-like motion, the plasma inside the jet has a perpendicular motion to the magnetic field lines, and the shape of these field lines changes accordingly, as shown by the blue curves in panels (a2) and (a3). To investigate the motion of the jet more clearly, we make two slices along the positions denoted by the dashed lines in panels (a4) and (a5), and the evolution of the number density along these two positions is shown in Figs.~\ref{Fig3}(b) and ~\ref{Fig3}(c), respectively. From panel (b), we can see that the emerging loop rises with a velocity of 10 km s$^{-1}$, in agreement with previous simulations and observations \citep{Chou1988, Shibata1989, Gonz2018, Tiwari2019}. 
After the reconnection, the cool materials are ejected upwards with a velocity of $\sim$ 55 km s$^{-1}$, which is comparable to the observed velocity of the jet. In panel (c), the upper blue dotted curve shows the whip-like motion of the jet, which in this time-distance view just appears as a damped transverse oscillation. The period of this oscillation is about 8 minutes. Not only does the magnetic field with the jet material have whip-like motions, but the whole loop system is oscillating vertically. As a result, condensations inside the underlying loops also displays forced oscillation, as shown by the lower dotted curve in panel (c).

Using the technique described in \citet{Xia2014}, we also obtain EUV synthetic images to compare the result with the AIA observations. Figure \ref{Fig4} demonstrates the morphology of the jet in 94 {\AA}, 171 {\AA}, 193 {\AA}, and 304 {\AA} wave channels at t = 236.151 min, which corresponds to plasma with temperatures of around 6.3, 0.8, 1.5, and 0.08 MK, respectively. In the hot 94 {\AA} image in panel (a), the hot X-ray jet could be observed as a bright feature, and the cool jet is dark. The cool jet also looks dark inside and bright at the boundary in the 171 {\AA} and 193 {\AA} channels. The positions near the reconnection area and the jet front are especially brighter in the 193 {\AA} channels, which is due to the heating caused by the reconnection and the shock. In the cool 304 {\AA} channel, the jet has a dark core and bright edges, which is an artifact of our 2.5D assumption since the boundary area of the jet has a TR temperature, which responds to the 304 {\AA} channels. We speculate that in a 3D simulation, the whole jet will look bright in 304 {\AA} channels.

As mentioned in Figure \ref{Fig3}, the movement of the plasmoid will change the shape of the local magnetic field lines, and we find that at scales accessible to current resolutions, this may trigger Kelvin--Helmholtz instability \citep[KHI; e.g.,][]{Keppens1999, Tian2016, NiL2017}. The KHI happens when two flows with strong velocity shears move along each other. In magnetized plasma, the occurrence of the KHI requires the total velocity difference to be more than a threshold, which is basically two times the Alfv\'en velocity, since a parallel magnetic field component can suppress KHI growth. As shown in Figures \ref{Fig5}(a1) and (a2), a heavy plasma blob drops from the top of the coronal loops. As the blob falls down, the magnetic field line becomes curved and stretched, so the weakened parallel magnetic fields can no longer suppress the KHI. This is shown by the Mach number maps in panels (b1) and (b2), where the Mach number at the interface is higher than 2, the threshold for the KHI. The deformed magnetic field is as shown by the black curves in panel (c). This falling plasma blob has a velocity difference with the ambient plasma, as seen by the vorticity map in panel (d). As the KHI happens, the boundary of the falling jet blobs display vortex-like structures shown in panel (e). There have been some observational works showing vortex-like structures at the edge of the jet \citep{LiX2018, LiX2019}, and we present a synthetic AIA 304 {\AA} image in panel (f) to compare with observations. The size of the vortex-like structures is approximately 1.5 Mm, which is consistent with the observations.

\section{Conclusion and discussions} \label{sec:conc}

In this letter, we report an event observed by \emph{SDO} and \emph{IRIS} where flux emergence at the northern end of a closed coronal loop system causes the eruption of plasma, forming a jet with an apparent velocity of 39 km s$^{-1}$ and a Doppler blue shift of 80 km s$^{-1}$. Adopting our open-source code MPI-AMRVAC, we perform a 2.5D MHD simulation which produces eruptions resulting from the flux emergence with a magnetic flux intensity of 150 G at the left footpoint of the loop. At first, the loops have coronal rain material inside them, and this coronal rain all moves to the other (right) side of the loops, since the emergence influences the pressure gradient inside the loops. Reconnection between the emerging flux and the preexisting loops forms a dense-plasma-loaded current sheet at the interface, and then magnetic islands and bi-directional flow appear. The magnetic islands coalesce into one blob and then are ejected outwards, which forms a cool jet that has a similar length, height, and velocity as the observations. The nearby plasma is heated to more than 4 MK, manifested as a hot jet. The hot jet could be observed as a bright feature in the synthetic 94 {\AA} images, and the cool jet is dark in the 94 {\AA}, 171 {\AA}, and 193 {\AA} channels. The jet exhibits distinct whip-like motions, and induces similar oscillations in the condensations in the loops beneath it. The movement of the jet material changes the local magnetic field configuration, which allows the occurrence of the KHI, and as a result vortex-like structures form at the edge of the jet.

Combined computational and observational efforts are needed to understand the interaction between flux emergence and dynamic atmospheric activities \citep{FanY2004, Cheung2014, Archontis2019}. We have successfully reproduced a lot of features reported in previous observations and simulations, such as magnetic islands, the simultaneous existence of hot and cool jets, whip-like motions, and vorticity structures due to KHI. Compared to the observed event, the jet we get has nearly the same properties (e.g., length, height, and velocities) as the observations. Our results confirm that magnetic reconnection between the emerging flux and overlying loops causes a series of coronal eruptions. There are a lot of parameters that could influence the trigger of the eruptions, such as the emerging magnetic flux strength ($B_1$), the relative angle (related to $\theta_1$ in our setup) between the emerging flux and the arcade \citep{Leake2022}. In follow-up work, we will present a detailed parameter survey to determine which parameters play the most prominent role.

Our simulation reveals that the influence of the moving multi-thermal plasma jet on the magnetic field can promote the KHI happening. Using a 2D MHD simulation, \citet{NiL2017} found that when the jet is ejected upwards, KHI could contribute to the formation of blob structures near the current sheet in a high $\beta$ case ($\sim$ 0.15 -- 1.5). \citet{LiX2018, LiX2019} suggested that KHI may be responsible for the vortex-like structures formed both during the upstream and downstream of the jets in the observations. In our simulation, the plasma $\beta$ around the place where the KHI happens is around 0.45. The movement of the jet materials could bend the magnetic field, making the local Mach number exceed the threshold of the KHI, resulting in vortex-like structures at the boundary of the jet. The scale of the vortex-like structures is similar to the observations. In the simulation, the high Mach number regions only exist for several minutes and break down soon after, leading to the suppression of the KHI, and then the vortex-like structures disappear. This explains why only a few KHI cases have been observed in jets so far, and we expect  that more KHI cases could be found in higher resolution observations. Also, it is important to revisit this setup to achieve a more comprehensive understanding of the system's behavior in full 3D settings.

\begin{acknowledgments}
We thank the referee for valuable suggestions. This work is done under the support of the European Research Council (ERC) under the European Unions Horizon 2020 research and innovation program (grant agreement No. 833251 PROMINENT ERC-ADG 2018). This work is further supported by an FWO project G0B4521N and the project C14/19/089 TRACESpace. YZ acknowledges funding from Research Foundation – Flanders FWO under project number 1256423N. The computational resources and services used in this work were provided by the VSC (Flemish Supercomputer Center), funded by the Research Foundation Flanders (FWO) and the Flemish Government - department EWI. The data used are courtesy of the \emph{SDO} and the \emph{IRIS} science teams.
\end{acknowledgments}

\software{\href{https://www.python.org/}{Python}, \href{https://yt-project.org/}{yt} \citep{Turk2011}}

\vspace{5mm}

\bibliography{sample631}{}

\begin{thebibliography}{}
\expandafter\ifx\csname natexlab\endcsname\relax\def\natexlab#1{#1}\fi
\providecommand{\url}[1]{\href{#1}{#1}}
\providecommand{\dodoi}[1]{doi:~\href{http://doi.org/#1}{\nolinkurl{#1}}}
\providecommand{\doeprint}[1]{\href{http://ascl.net/#1}{\nolinkurl{http://ascl.net/#1}}}
\providecommand{\doarXiv}[1]{\href{https://arxiv.org/abs/#1}{\nolinkurl{https://arxiv.org/abs/#1}}}

\bibitem[{{Archontis} {et~al.}(2004){Archontis}, {Moreno-Insertis},
  {Galsgaard}, {Hood}, \& {O'Shea}}]{Archontis2004}
{Archontis}, V., {Moreno-Insertis}, F., {Galsgaard}, K., {Hood}, A., \&
  {O'Shea}, E. 2004, \aap, 426, 1047, \dodoi{10.1051/0004-6361:20035934}

\bibitem[{{Archontis} \& {Syntelis}(2019)}]{Archontis2019}
{Archontis}, V., \& {Syntelis}, P. 2019, Philosophical Transactions of the
  Royal Society of London Series A, 377, 20180387,
  \dodoi{10.1098/rsta.2018.0387}

\bibitem[{{Caligari} {et~al.}(1995){Caligari}, {Moreno-Insertis}, \&
  {Schussler}}]{Caligari1995}
{Caligari}, P., {Moreno-Insertis}, F., \& {Schussler}, M. 1995, \apj, 441, 886,
  \dodoi{10.1086/175410}

\bibitem[{{Canfield} {et~al.}(1996){Canfield}, {Reardon}, {Leka}, {Shibata},
  {Yokoyama}, \& {Shimojo}}]{Canfield1996}
{Canfield}, R.~C., {Reardon}, K.~P., {Leka}, K.~D., {et~al.} 1996, \apj, 464,
  1016, \dodoi{10.1086/177389}

\bibitem[{{Chandrashekhar} {et~al.}(2014){Chandrashekhar}, {Bemporad},
  {Banerjee}, {Gupta}, \& {Teriaca}}]{Chandrashekhar2014}
{Chandrashekhar}, K., {Bemporad}, A., {Banerjee}, D., {Gupta}, G.~R., \&
  {Teriaca}, L. 2014, \aap, 561, A104, \dodoi{10.1051/0004-6361/201321213}

\bibitem[{{Chen} {et~al.}(2017){Chen}, {Rempel}, \& {Fan}}]{ChenF2017}
{Chen}, F., {Rempel}, M., \& {Fan}, Y. 2017, \apj, 846, 149,
  \dodoi{10.3847/1538-4357/aa85a0}

\bibitem[{{Chen}(2011)}]{ChenP2011}
{Chen}, P.~F. 2011, Living Reviews in Solar Physics, 8, 1,
  \dodoi{10.12942/lrsp-2011-1}

\bibitem[{{Chen} \& {Shibata}(2000)}]{ChenP2000}
{Chen}, P.~F., \& {Shibata}, K. 2000, \apj, 545, 524, \dodoi{10.1086/317803}

\bibitem[{{Cheung} \& {Isobe}(2014)}]{Cheung2014}
{Cheung}, M. C.~M., \& {Isobe}, H. 2014, Living Reviews in Solar Physics, 11,
  3, \dodoi{10.12942/lrsp-2014-3}

\bibitem[{{Cheung} {et~al.}(2010){Cheung}, {Rempel}, {Title}, \&
  {Sch{\"u}ssler}}]{Cheung2010}
{Cheung}, M.~C.~M., {Rempel}, M., {Title}, A.~M., \& {Sch{\"u}ssler}, M. 2010,
  \apj, 720, 233, \dodoi{10.1088/0004-637X/720/1/233}

\bibitem[{{Cheung} {et~al.}(2019){Cheung}, {Rempel}, {Chintzoglou}, {Chen},
  {Testa}, {Mart{\'\i}nez-Sykora}, {Sainz Dalda}, {DeRosa}, {Malanushenko},
  {Hansteen}, {De Pontieu}, {Carlsson}, {Gudiksen}, \& {McIntosh}}]{Cheung2019}
{Cheung}, M.~C.~M., {Rempel}, M., {Chintzoglou}, G., {et~al.} 2019, Nature
  Astronomy, 3, 160, \dodoi{10.1038/s41550-018-0629-3}

\bibitem[{{Chou} \& {Zirin}(1988)}]{Chou1988}
{Chou}, D.-Y., \& {Zirin}, H. 1988, \apj, 333, 420, \dodoi{10.1086/166757}

\bibitem[{{De Pontieu} {et~al.}(2014){De Pontieu}, {Title}, {Lemen}, {Kushner},
  {Akin}, {Allard}, {Berger}, {Boerner}, {Cheung}, {Chou}, {Drake}, {Duncan},
  {Freeland}, {Heyman}, {Hoffman}, {Hurlburt}, {Lindgren}, {Mathur}, {Rehse},
  {Sabolish}, {Seguin}, {Schrijver}, {Tarbell}, {W{\"u}lser}, {Wolfson},
  {Yanari}, {Mudge}, {Nguyen-Phuc}, {Timmons}, {van Bezooijen}, {Weingrod},
  {Brookner}, {Butcher}, {Dougherty}, {Eder}, {Knagenhjelm}, {Larsen},
  {Mansir}, {Phan}, {Boyle}, {Cheimets}, {DeLuca}, {Golub}, {Gates}, {Hertz},
  {McKillop}, {Park}, {Perry}, {Podgorski}, {Reeves}, {Saar}, {Testa}, {Tian},
  {Weber}, {Dunn}, {Eccles}, {Jaeggli}, {Kankelborg}, {Mashburn}, {Pust},
  {Springer}, {Carvalho}, {Kleint}, {Marmie}, {Mazmanian}, {Pereira}, {Sawyer},
  {Strong}, {Worden}, {Carlsson}, {Hansteen}, {Leenaarts}, {Wiesmann},
  {Aloise}, {Chu}, {Bush}, {Scherrer}, {Brekke}, {Martinez-Sykora}, {Lites},
  {McIntosh}, {Uitenbroek}, {Okamoto}, {Gummin}, {Auker}, {Jerram}, {Pool}, \&
  {Waltham}}]{DeP2014}
{De Pontieu}, B., {Title}, A.~M., {Lemen}, J.~R., {et~al.} 2014, \solphys, 289,
  2733, \dodoi{10.1007/s11207-014-0485-y}

\bibitem[{{Fan}(2001)}]{FanY2001}
{Fan}, Y. 2001, \apjl, 554, L111, \dodoi{10.1086/320935}

\bibitem[{{Fan}(2021)}]{FanY2021}
---. 2021, Living Reviews in Solar Physics, 18, 5,
  \dodoi{10.1007/s41116-021-00031-2}

\bibitem[{{Fan} \& {Gibson}(2004)}]{FanY2004}
{Fan}, Y., \& {Gibson}, S.~E. 2004, \apj, 609, 1123, \dodoi{10.1086/421238}

\bibitem[{{Fang} {et~al.}(2013){Fang}, {Xia}, \& {Keppens}}]{Fang2013}
{Fang}, X., {Xia}, C., \& {Keppens}, R. 2013, \apjl, 771, L29,
  \dodoi{10.1088/2041-8205/771/2/L29}

\bibitem[{{Fang} {et~al.}(2015){Fang}, {Xia}, {Keppens}, \& {Van
  Doorsselaere}}]{Fang2015}
{Fang}, X., {Xia}, C., {Keppens}, R., \& {Van Doorsselaere}, T. 2015, \apj,
  807, 142, \dodoi{10.1088/0004-637X/807/2/142}

\bibitem[{{Feynman} \& {Martin}(1995)}]{Feynman1995}
{Feynman}, J., \& {Martin}, S.~F. 1995, \jgr, 100, 3355,
  \dodoi{10.1029/94JA02591}

\bibitem[{{Forbes} \& {Priest}(1983)}]{Forbes1983}
{Forbes}, T.~G., \& {Priest}, E.~R. 1983, \solphys, 84, 169,
  \dodoi{10.1007/BF00157455}

\bibitem[{{Gonz{\'a}lez Manrique} {et~al.}(2018){Gonz{\'a}lez Manrique},
  {Kuckein}, {Collados}, {Denker}, {Solanki}, {G{\"o}m{\"o}ry}, {Verma},
  {Balthasar}, {Lagg}, \& {Diercke}}]{Gonz2018}
{Gonz{\'a}lez Manrique}, S.~J., {Kuckein}, C., {Collados}, M., {et~al.} 2018,
  \aap, 617, A55, \dodoi{10.1051/0004-6361/201832684}

\bibitem[{{Guglielmino} {et~al.}(2010){Guglielmino}, {Bellot Rubio},
  {Zuccarello}, {Aulanier}, {Vargas Dom{\'\i}nguez}, \&
  {Kamio}}]{Guglielmino2010}
{Guglielmino}, S.~L., {Bellot Rubio}, L.~R., {Zuccarello}, F., {et~al.} 2010,
  \apj, 724, 1083, \dodoi{10.1088/0004-637X/724/2/1083}

\bibitem[{{Guglielmino} {et~al.}(2018){Guglielmino}, {Zuccarello}, {Young},
  {Murabito}, \& {Romano}}]{Guglielmino2018}
{Guglielmino}, S.~L., {Zuccarello}, F., {Young}, P.~R., {Murabito}, M., \&
  {Romano}, P. 2018, \apj, 856, 127, \dodoi{10.3847/1538-4357/aab2a8}

\bibitem[{{Hansteen} {et~al.}(2019){Hansteen}, {Ortiz}, {Archontis},
  {Carlsson}, {Pereira}, \& {Bj{\o}rgen}}]{Hansteen2019}
{Hansteen}, V., {Ortiz}, A., {Archontis}, V., {et~al.} 2019, \aap, 626, A33,
  \dodoi{10.1051/0004-6361/201935376}

\bibitem[{{Heyvaerts} {et~al.}(1977){Heyvaerts}, {Priest}, \&
  {Rust}}]{Heyvaerts1977}
{Heyvaerts}, J., {Priest}, E.~R., \& {Rust}, D.~M. 1977, \apj, 216, 123,
  \dodoi{10.1086/155453}

\bibitem[{{Innes} {et~al.}(1997){Innes}, {Inhester}, {Axford}, \&
  {Wilhelm}}]{Innes1997}
{Innes}, D.~E., {Inhester}, B., {Axford}, W.~I., \& {Wilhelm}, K. 1997, \nat,
  386, 811, \dodoi{10.1038/386811a0}

\bibitem[{{Jiang} {et~al.}(2016){Jiang}, {Wu}, {Feng}, \& {Hu}}]{JiangC2016}
{Jiang}, C., {Wu}, S.~T., {Feng}, X., \& {Hu}, Q. 2016, Nature Communications,
  7, 11522, \dodoi{10.1038/ncomms11522}

\bibitem[{{Johnston} \& {Bradshaw}(2019)}]{Johnston2019}
{Johnston}, C.~D., \& {Bradshaw}, S.~J. 2019, \apjl, 873, L22,
  \dodoi{10.3847/2041-8213/ab0c1f}

\bibitem[{{Keppens} {et~al.}(2012){Keppens}, {Meliani}, {van Marle}, {Delmont},
  {Vlasis}, \& {van der Holst}}]{Keppens2012}
{Keppens}, R., {Meliani}, Z., {van Marle}, A.~J., {et~al.} 2012, Journal of
  Computational Physics, 231, 718, \dodoi{10.1016/j.jcp.2011.01.020}

\bibitem[{{Keppens} {et~al.}(2021){Keppens}, {Teunissen}, {Xia}, \&
  {Porth}}]{Keppens2021}
{Keppens}, R., {Teunissen}, J., {Xia}, C., \& {Porth}, O. 2021, Computers and
  Mathematics with Applications, 81, 316,
  \dodoi{https://doi.org/10.1016/j.camwa.2020.03.023}

\bibitem[{{Keppens} {et~al.}(1999){Keppens}, {T{\'o}th}, {Westermann}, \&
  {Goedbloed}}]{Keppens1999}
{Keppens}, R., {T{\'o}th}, G., {Westermann}, R.~H.~J., \& {Goedbloed}, J.~P.
  1999, Journal of Plasma Physics, 61, 1, \dodoi{10.1017/S0022377898007223}

\bibitem[{{Kliem} \& {T{\"o}r{\"o}k}(2006)}]{Kliem2006}
{Kliem}, B., \& {T{\"o}r{\"o}k}, T. 2006, \prl, 96, 255002,
  \dodoi{10.1103/PhysRevLett.96.255002}

\bibitem[{{Leake} {et~al.}(2014){Leake}, {Linton}, \& {Antiochos}}]{leake2014}
{Leake}, J.~E., {Linton}, M.~G., \& {Antiochos}, S.~K. 2014, \apj, 787, 46,
  \dodoi{10.1088/0004-637X/787/1/46}

\bibitem[{{Leake} {et~al.}(2022){Leake}, {Linton}, \& {Antiochos}}]{Leake2022}
---. 2022, \apj, 934, 10, \dodoi{10.3847/1538-4357/ac74b7}

\bibitem[{{Leake} {et~al.}(2013){Leake}, {Linton}, \&
  {T{\"o}r{\"o}k}}]{Leake2013}
{Leake}, J.~E., {Linton}, M.~G., \& {T{\"o}r{\"o}k}, T. 2013, \apj, 778, 99,
  \dodoi{10.1088/0004-637X/778/2/99}

\bibitem[{{Lemen} {et~al.}(2012){Lemen}, {Title}, {Akin}, {Boerner}, {Chou},
  {Drake}, {Duncan}, {Edwards}, {Friedlaender}, {Heyman}, {Hurlburt}, {Katz},
  {Kushner}, {Levay}, {Lindgren}, {Mathur}, {McFeaters}, {Mitchell}, {Rehse},
  {Schrijver}, {Springer}, {Stern}, {Tarbell}, {Wuelser}, {Wolfson}, {Yanari},
  {Bookbinder}, {Cheimets}, {Caldwell}, {Deluca}, {Gates}, {Golub}, {Park},
  {Podgorski}, {Bush}, {Scherrer}, {Gummin}, {Smith}, {Auker}, {Jerram},
  {Pool}, {Soufli}, {Windt}, {Beardsley}, {Clapp}, {Lang}, \&
  {Waltham}}]{Lemen2012}
{Lemen}, J.~R., {Title}, A.~M., {Akin}, D.~J., {et~al.} 2012, \solphys, 275,
  17, \dodoi{10.1007/s11207-011-9776-8}

\bibitem[{{Li} {et~al.}(2022){Li}, {Keppens}, \& {Zhou}}]{LiX2022}
{Li}, X., {Keppens}, R., \& {Zhou}, Y. 2022, \apj, 926, 216,
  \dodoi{10.3847/1538-4357/ac41cd}

\bibitem[{{Li} {et~al.}(2019){Li}, {Zhang}, {Yang}, \& {Hou}}]{LiX2019}
{Li}, X., {Zhang}, J., {Yang}, S., \& {Hou}, Y. 2019, \apj, 875, 52,
  \dodoi{10.3847/1538-4357/ab0f39}

\bibitem[{{Li} {et~al.}(2018){Li}, {Zhang}, {Yang}, {Hou}, \&
  {Erd{\'e}lyi}}]{LiX2018}
{Li}, X., {Zhang}, J., {Yang}, S., {Hou}, Y., \& {Erd{\'e}lyi}, R. 2018,
  Scientific Reports, 8, 8136, \dodoi{10.1038/s41598-018-26581-4}

\bibitem[{{Lionello} {et~al.}(2016){Lionello}, {T{\"o}r{\"o}k}, {Titov},
  {Leake}, {Miki{\'c}}, {Linker}, \& {Linton}}]{Lionello2016}
{Lionello}, R., {T{\"o}r{\"o}k}, T., {Titov}, V.~S., {et~al.} 2016, \apjl, 831,
  L2, \dodoi{10.3847/2041-8205/831/1/L2}

\bibitem[{{Manchester} {et~al.}(2004){Manchester}, {Gombosi}, {DeZeeuw}, \&
  {Fan}}]{Manchester2004}
{Manchester}, W., I., {Gombosi}, T., {DeZeeuw}, D., \& {Fan}, Y. 2004, \apj,
  610, 588, \dodoi{10.1086/421516}

\bibitem[{{Meyer} {et~al.}(2012){Meyer}, {Balsara}, \& {Aslam}}]{Meyer2012}
{Meyer}, C.~D., {Balsara}, D.~S., \& {Aslam}, T.~D. 2012, \mnras, 422, 2102,
  \dodoi{10.1111/j.1365-2966.2012.20744.x}

\bibitem[{{Meyer} {et~al.}(2014){Meyer}, {Balsara}, \& {Aslam}}]{Meyer2014}
---. 2014, Journal of Computational Physics, 257, 594,
  \dodoi{10.1016/j.jcp.2013.08.021}

\bibitem[{{Miyagoshi} \& {Yokoyama}(2004)}]{Miyagoshi2004}
{Miyagoshi}, T., \& {Yokoyama}, T. 2004, \apj, 614, 1042,
  \dodoi{10.1086/423731}

\bibitem[{{Moreno-Insertis} {et~al.}(2008){Moreno-Insertis}, {Galsgaard}, \&
  {Ugarte-Urra}}]{Moreno2008}
{Moreno-Insertis}, F., {Galsgaard}, K., \& {Ugarte-Urra}, I. 2008, \apjl, 673,
  L211, \dodoi{10.1086/527560}

\bibitem[{{Moreno-Insertis} {et~al.}(2018){Moreno-Insertis}, {Martinez-Sykora},
  {Hansteen}, \& {Mu{\~n}oz}}]{Moreno2018}
{Moreno-Insertis}, F., {Martinez-Sykora}, J., {Hansteen}, V.~H., \&
  {Mu{\~n}oz}, D. 2018, \apjl, 859, L26, \dodoi{10.3847/2041-8213/aac648}

\bibitem[{{Morton} {et~al.}(2012){Morton}, {Srivastava}, \&
  {Erd{\'e}lyi}}]{Morton2012}
{Morton}, R.~J., {Srivastava}, A.~K., \& {Erd{\'e}lyi}, R. 2012, \aap, 542,
  A70, \dodoi{10.1051/0004-6361/201117218}

\bibitem[{{Ni} {et~al.}(2017){Ni}, {Zhang}, {Murphy}, \& {Lin}}]{NiL2017}
{Ni}, L., {Zhang}, Q.-M., {Murphy}, N.~A., \& {Lin}, J. 2017, \apj, 841, 27,
  \dodoi{10.3847/1538-4357/aa6ffe}

\bibitem[{{N{\'o}brega-Siverio} \& {Moreno-Insertis}(2022)}]{Nobrega2022}
{N{\'o}brega-Siverio}, D., \& {Moreno-Insertis}, F. 2022, \apjl, 935, L21,
  \dodoi{10.3847/2041-8213/ac85b6}

\bibitem[{{N{\'o}brega-Siverio} {et~al.}(2016){N{\'o}brega-Siverio},
  {Moreno-Insertis}, \& {Mart{\'\i}nez-Sykora}}]{Nobrega2016}
{N{\'o}brega-Siverio}, D., {Moreno-Insertis}, F., \& {Mart{\'\i}nez-Sykora}, J.
  2016, \apj, 822, 18, \dodoi{10.3847/0004-637X/822/1/18}

\bibitem[{{Pesnell} {et~al.}(2012){Pesnell}, {Thompson}, \&
  {Chamberlin}}]{Pesnell2012}
{Pesnell}, W.~D., {Thompson}, B.~J., \& {Chamberlin}, P.~C. 2012, \solphys,
  275, 3, \dodoi{10.1007/s11207-011-9841-3}

\bibitem[{{Pontin} \& {Priest}(2022)}]{Pontin2022}
{Pontin}, D.~I., \& {Priest}, E.~R. 2022, Living Reviews in Solar Physics, 19,
  1, \dodoi{10.1007/s41116-022-00032-9}

\bibitem[{{Porth} {et~al.}(2014){Porth}, {Xia}, {Hendrix}, {Moschou}, \&
  {Keppens}}]{Porth2014}
{Porth}, O., {Xia}, C., {Hendrix}, T., {Moschou}, S.~P., \& {Keppens}, R. 2014,
  \apjs, 214, 4, \dodoi{10.1088/0067-0049/214/1/4}

\bibitem[{{Schmieder} {et~al.}(2014){Schmieder}, {Archontis}, \&
  {Pariat}}]{Schmieder2014}
{Schmieder}, B., {Archontis}, V., \& {Pariat}, E. 2014, \ssr, 186, 227,
  \dodoi{10.1007/s11214-014-0088-9}

\bibitem[{{Schrijver}(2009)}]{Schrijver2009}
{Schrijver}, C.~J. 2009, Advances in Space Research, 43, 739,
  \dodoi{10.1016/j.asr.2008.11.004}

\bibitem[{{Shen} {et~al.}(2022){Shen}, {Xu}, {Li}, \& {Ji}}]{Shen2022}
{Shen}, J., {Xu}, Z., {Li}, J., \& {Ji}, H. 2022, \apj, 925, 46,
  \dodoi{10.3847/1538-4357/ac37c3}

\bibitem[{{Shibata} {et~al.}(1992){Shibata}, {Nozawa}, \&
  {Matsumoto}}]{Shibata1992}
{Shibata}, K., {Nozawa}, S., \& {Matsumoto}, R. 1992, \pasj, 44, 265

\bibitem[{{Shibata} {et~al.}(1989){Shibata}, {Tajima}, {Steinolfson}, \&
  {Matsumoto}}]{Shibata1989}
{Shibata}, K., {Tajima}, T., {Steinolfson}, R.~S., \& {Matsumoto}, R. 1989,
  \apj, 345, 584, \dodoi{10.1086/167932}

\bibitem[{{Shimojo} {et~al.}(2001){Shimojo}, {Shibata}, {Yokoyama}, \&
  {Hori}}]{Shimojo2001}
{Shimojo}, M., {Shibata}, K., {Yokoyama}, T., \& {Hori}, K. 2001, \apj, 550,
  1051, \dodoi{10.1086/319788}

\bibitem[{{Tian} \& {Chen}(2016)}]{Tian2016}
{Tian}, C., \& {Chen}, Y. 2016, \apj, 824, 60,
  \dodoi{10.3847/0004-637X/824/1/60}

\bibitem[{{Tiwari} {et~al.}(2019){Tiwari}, {Panesar}, {Moore}, {De Pontieu},
  {Winebarger}, {Golub}, {Savage}, {Rachmeler}, {Kobayashi}, {Testa}, {Warren},
  {Brooks}, {Cirtain}, {McKenzie}, {Morton}, {Peter}, \& {Walsh}}]{Tiwari2019}
{Tiwari}, S.~K., {Panesar}, N.~K., {Moore}, R.~L., {et~al.} 2019, \apj, 887,
  56, \dodoi{10.3847/1538-4357/ab54c1}

\bibitem[{{Toriumi}(2022)}]{Toriumi2022}
{Toriumi}, S. 2022, Advances in Space Research, 70, 1549,
  \dodoi{10.1016/j.asr.2021.05.017}

\bibitem[{{Toriumi} \& {Wang}(2019)}]{Toriumi2019}
{Toriumi}, S., \& {Wang}, H. 2019, Living Reviews in Solar Physics, 16, 3,
  \dodoi{10.1007/s41116-019-0019-7}

\bibitem[{{T{\"o}r{\"o}k} {et~al.}(2016){T{\"o}r{\"o}k}, {Lionello}, {Titov},
  {Leake}, {Miki{\'c}}, {Linker}, \& {Linton}}]{Torok2016}
{T{\"o}r{\"o}k}, T., {Lionello}, R., {Titov}, V.~S., {et~al.} 2016, in
  Astronomical Society of the Pacific Conference Series, Vol. 504, Coimbra
  Solar Physics Meeting: Ground-based Solar Observations in the Space
  Instrumentation Era, ed. I.~{Dorotovic}, C.~E. {Fischer}, \& M.~{Temmer}, 185

\bibitem[{{Townsend}(2009)}]{Townsend2009}
{Townsend}, R.~H.~D. 2009, \apjs, 181, 391, \dodoi{10.1088/0067-0049/181/2/391}

\bibitem[{{Turk} {et~al.}(2011){Turk}, {Smith}, {Oishi}, {Skory}, {Skillman},
  {Abel}, \& {Norman}}]{Turk2011}
{Turk}, M.~J., {Smith}, B.~D., {Oishi}, J.~S., {et~al.} 2011, \apjs, 192, 9,
  \dodoi{10.1088/0067-0049/192/1/9}

\bibitem[{{van Driel-Gesztelyi} \& {Green}(2015)}]{vanD2015}
{van Driel-Gesztelyi}, L., \& {Green}, L.~M. 2015, Living Reviews in Solar
  Physics, 12, 1, \dodoi{10.1007/lrsp-2015-1}

\bibitem[{{Wang} {et~al.}(2020){Wang}, {Liu}, {Cao}, \& {Wang}}]{Wang2020}
{Wang}, J., {Liu}, C., {Cao}, W., \& {Wang}, H. 2020, \apj, 900, 84,
  \dodoi{10.3847/1538-4357/aba696}

\bibitem[{{Xia} {et~al.}(2014){Xia}, {Keppens}, {Antolin}, \&
  {Porth}}]{Xia2014}
{Xia}, C., {Keppens}, R., {Antolin}, P., \& {Porth}, O. 2014, \apjl, 792, L38,
  \dodoi{10.1088/2041-8205/792/2/L38}

\bibitem[{{Xia} {et~al.}(2018){Xia}, {Teunissen}, {El Mellah}, {Chan{\'e}}, \&
  {Keppens}}]{Xia2018}
{Xia}, C., {Teunissen}, J., {El Mellah}, I., {Chan{\'e}}, E., \& {Keppens}, R.
  2018, \apjs, 234, 30, \dodoi{10.3847/1538-4365/aaa6c8}

\bibitem[{{Yokoyama} \& {Shibata}(1995)}]{Yokoyama1995}
{Yokoyama}, T., \& {Shibata}, K. 1995, \nat, 375, 42, \dodoi{10.1038/375042a0}

\bibitem[{{Yokoyama} \& {Shibata}(1996)}]{Yokoyama1996}
---. 1996, \pasj, 48, 353, \dodoi{10.1093/pasj/48.2.353}

\bibitem[{{Zhou} {et~al.}(2020){Zhou}, {Chen}, {Hong}, \& {Fang}}]{Zhou2020}
{Zhou}, Y.~H., {Chen}, P.~F., {Hong}, J., \& {Fang}, C. 2020, Nature Astronomy,
  4, 994, \dodoi{10.1038/s41550-020-1094-3}

\bibitem[{{Zhou} {et~al.}(2021){Zhou}, {Ruan}, {Xia}, \& {Keppens}}]{ZhouY2021}
{Zhou}, Y.-H., {Ruan}, W.-Z., {Xia}, C., \& {Keppens}, R. 2021, \aap, 648, A29,
  \dodoi{10.1051/0004-6361/202040254}

\bibitem[{{Zwaan}(1985)}]{Zwaan1985}
{Zwaan}, C. 1985, \solphys, 100, 397, \dodoi{10.1007/BF00158438}

\bibitem[{{Zwaan}(1987)}]{Zwaan1987}
---. 1987, \araa, 25, 83, \dodoi{10.1146/annurev.aa.25.090187.000503}

\end{thebibliography}
\bibliographystyle{aasjournal}

\end{document}